\documentclass[aip,reprint]{revtex4-1}

\usepackage{graphicx}
\usepackage{dcolumn}
\usepackage{bm}
\usepackage[mathlines]{lineno}
\usepackage[utf8]{inputenc}
\usepackage[T1]{fontenc}
\usepackage{mathptmx}
\usepackage{xcolor}
\usepackage{gensymb}
\usepackage{braket}
\makeatletter
\newcommand\footnoteref[1]{\protected@xdef\@thefnmark{\ref{#1}}\@footnotemark}
\makeatother
\draft
\begin{document}

\title[]{Large Microwave Inductance of Granular Boron-Doped Diamond Superconducting Films}
\author{Bakhrom Oripov}
\email{bakhromtjk@gmail.com}
\affiliation{Quantum Materials Center, Physics Department,\\University of Maryland, College Park, MD 20742, USA}

\author{Dinesh Kumar}
\affiliation{Department of Physics, Quantum Centres in Diamond and Emergent Materials (QuCenDiEM)-group, Nano Functional Materials Technology Centre and Materials Science Research Centre, Indian Institute of Technology Madras, Chennai 600036, India}

\author{Cougar Garcia}
\affiliation{Quantum Materials Center, Physics Department,\\University of Maryland, College Park, MD 20742, USA}
\affiliation{Materials Science and Engineering Department,\\University of Maryland, College Park, MD 20742, USA}

\author{Patrick Hemmer}
\affiliation{Quantum Materials Center, Physics Department,\\University of Maryland, College Park, MD 20742, USA}

\author{T. Venkatesan}
\affiliation{NUSNNI NanoCore, National University of Singapore, 117411, Singapore}
\altaffiliation[Current Address:]{ Center for Quantum Research and Technology, The University of Oklahoma, 440 W. Brooks Street, Norman, Oklahoma 73019, USA}
\affiliation{Neocera LLC, Beltsville, MD 20705, USA}

\author{M. S. Ramachandra Rao}
\affiliation{Department of Physics, Quantum Centres in Diamond and Emergent Materials (QuCenDiEM)-group, Nano Functional Materials Technology Centre and Materials Science Research Centre, Indian Institute of Technology Madras, Chennai 600036, India}

\author{Steven M. Anlage}
\affiliation{Quantum Materials Center, Physics Department,\\University of Maryland, College Park, MD 20742, USA}
\affiliation{Materials Science and Engineering Department,\\University of Maryland, College Park, MD 20742, USA}

\date{\today}

\begin{abstract}
Boron-doped diamond granular thin films are known to exhibit superconductivity with an optimal critical temperature of $T_c = 7.2K$. Here we report the measured in-plane complex surface impedance of Boron-doped diamond films in the microwave frequency range using a resonant technique.  Experimentally measured inductance values are in good agreement with estimates obtained from the normal state sheet resistance of the material.  The magnetic penetration depth temperature dependence is consistent with that of a fully-gapped s-wave superconductor.  Boron-doped diamond films should find application where high kinetic inductance is needed, such as microwave kinetic inductance detectors and quantum impedance devices.
\end{abstract}

\maketitle
Diamond is the hardest known natural material and an excellent thermal conductor despite also being electrically insulating with a large band gap. Doping diamond can create either semiconducting or metallic states.\cite{Yokoya2005,Watanabe2012} It was discovered from high pressure and high temperature (HPHT) synthesis techniques that bulk single crystal diamond can be doped with Boron (hole doping) to the point that superconductivity is observed with $T_c\cong2.3\;K$.\cite{Sidorov2010} The first thin films of Boron-doped diamond (BDD) were single-crystal like and had a transition temperature that increased with Boron doping, but only slightly exceeded that obtained in bulk.\cite{Bustarret2004} Later it was discovered that higher B concentrations could be obtained by preparing the films with energetic and non-equilibrium methods, including microwave plasma-assisted chemical vapor deposition (MPCVD),\cite{Takano2004} which resulted in $T_c\cong4.2\;K$. Upon further refinement, materials that have become known as nano-crystalline diamond (NCD) prepared by MPCVD showed even higher transition temperatures. A remarkable dome-shaped $T_c$ vs. doping curve has emerged from extensive studies of boron doped granular diamond films, with a peak transition temperature of $T_c\cong7.2\;K$.\cite{Kumar2018}  Meanwhile Moussa and Cohen have predicted that the transition temperature of B-doped diamond will grow to 55 K with B concentration in the range of 20\% to 30\%.\cite{Moussa08}  There are also predictions of high transition temperature superconductivity in metastable crystals of Boron-Carbon compounds based on electronic and vibrational numerical calculations.\cite{Saha20}  Recently superconductivity with a zero-resistance $T_c=$ 24 K has been discovered in 27 at\% B-doped rapidly quenched Carbon.\cite{Bha17}  Meanwhile, BDD films have also found a broad set of uses in sensing,\cite{Ensch18} electroanalytics,\cite{Muz19} catalysis,\cite{BinLiu20}, semiconductor devices,\cite{Yao21} and as conducting contacts in various applications.\cite{Ficek16,Man21}
\par

The single-crystal-like B-doped diamond films show clear signs of BCS s-wave superconducting behavior in STM tunneling experiments, with no sub-gap states and a peak in the density of states consistent with $\Delta(0)/k_BT_c=1.74\;$.\cite{Sacepe2006} However it was noted from scanning tunneling spectroscopy methods that Boron-doped diamond NCD films displayed modulation of the superconducting order parameter between grains, strong superconducting fluctuations, and substantial tunneling between grains in the superconducting state, all consistent with a strong granular nature to superconductivity.\cite{Willems2010,Klem21} It was noted that diamond grains with different exposed surfaces may absorb Boron to a widely varying extent, resulting in locally inhomogeneous superconducting properties in NCD films.\cite{Dahlem2010,Zhang2011} It was argued that doped NCD films are near a superconductor-to-insulator transition driven by competition between Coulomb blockade in the grains and the superconducting proximity effect that encourages long-range transport.  The NCD films have residual resistance ratio values less than unity \cite{Zhang2011} and show large values of resistivity ($\rho_n=2-40\;m\Omega\;cm$) just above the transition temperature.\cite{Dahlem2010,Kumar2018}  As one consequence of this complex microstructure, the doped NCD films show local values of $\Delta(0)/k_BT_c$ substantially less than the weak-coupling limit value of 1.76.\cite{Dahlem2010} \par

For some time it has been known that the kinetic inductance of superconductors in narrow thin film wires can exceed the geometrical inductance of the structure.\cite{Little1967,Meservey1969} The use of low carrier density materials, disordered superconductors, and granular materials has further enhanced the effective inductance of patterned film structures. These high effective inductance materials have enabled compact microwave resonators,\cite{Zhang2019a} and a new generation of highly sensitive photon detectors based on lumped-element microwave resonators \cite{Zmuidzinas2012}. They also find use as high-impedance (exceeding $R_Q=h/(2e)^2\simeq6.5\;k\Omega$) structures for quantum circuits, current-tunable delay lines \cite{Anlage1989, Annunziata2010}, phase shifters,\cite{Bockstiegel2016} etc. Superconducting materials of interest include NbN$_x$ \cite{Peltonen2013}, InO$_x$,\cite{Fiory1983} NbTiN,\cite{Kawamura1999} TiN,\cite{Leduc2010} and granular Aluminum (grAl).\cite{Cohen1968a, Maleeva2018} For a BCS superconductor in the dirty limit, one can relate the superconducting inductance to the normal state sheet resistance $R_n$ and superconducting gap $\Delta$ as $L_s=\hbar R_n/(\pi\Delta)$, with $R_n=\rho_n/t$, where $t<\lambda$ is the film thickness and $\lambda$ is the magnetic penetration depth.\cite{Zmuidzinas2012} Hence there has been a search for new materials with large normal state sheet resistances that do not compromise the superconducting properties (in terms of $T_c$ and $\Delta$) so that the material will retain the low microwave loss properties ($\sigma_1\ll\sigma_2$) associated with robust superconductors. \par

\begin{figure}[b!]
\includegraphics[width=0.5\textwidth]{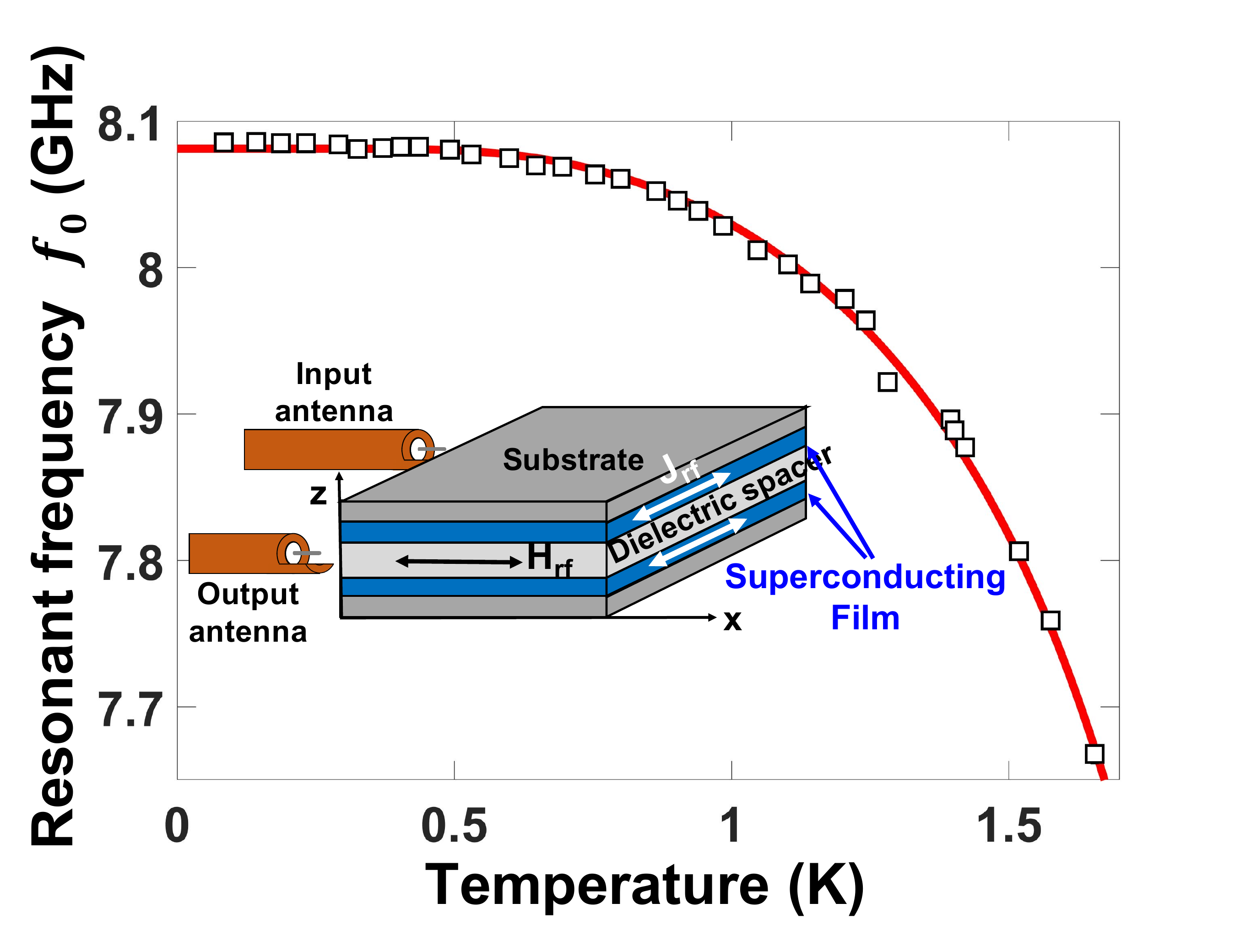}
\caption{ Temperature dependence of the resonant frequency $f_{0}(T)$ extracted from PPR measurement on Boron-doped diamond film "B" (black squares) and fits to the data using Eq.(\ref{Eq_PPR_ft}) (solid red line). Here the error bars are smaller than the symbol size. Inset shows a schematic of the Parallel Plate Resonator and coaxial microwave cables with capacitive coupling.  The directions of rf magnetic field and currents for a representative mode are shown.  Inset figure not to scale.  \label{Fig_SampleB}}
\end{figure}

We utilize the parallel plate resonator (PPR) technique to directly measure the superconducting inductance and microwave losses of a new class of granular materials, namely granular Boron-doped diamond films. In this approach, a solid low-loss dielectric spacer is placed between two flat and nominally identical superconducting films forming a parallel plate resonator, as shown in the inset of Fig. \ref{Fig_SampleB}.  Electromagnetic waves are launched into the open-sided structure making it into a quasi-two-dimensional resonator for the lowest order modes.  The resonance frequency is extremely sensitive to the superconducting magnetic penetration depth,\cite{Anlage1989} while the quality factor is sensitive to the microwave losses in the superconducting films.\cite{Taber1990}  No direct electrical contacts are made to the superconducting samples, preserving the open-circuit boundary conditions at the edges. Instead, two antennas are capacitively coupled to the PPR with a variable coupling strength that be tuned by changing the physical distance between the antennas and the PPR edge. The antennas are made by exposing a short part of the inner conductor and a portion of the outer conductor of a coaxial cable. Assuming that a thin enough dielectric spacer is used, the effects of the edge currents and imperfections in the shape or size of the sample can be safely ignored. \par

The resonant frequency of this resonator will be affected by the field penetration into the superconducting plates. The measured temperature dependence of the resonant frequency $f_{0}$ stems from the temperature dependence of the superconducting penetration depth $\lambda(T)$, and can be written as,\cite{Taber1990,Talanov2000}
\begin{equation}
    \frac{f_{0}(T)}{f_{0}(0)}=\frac{\left[1+2\frac{\lambda(0)}{d}coth\left(\frac{t}{\lambda(0)}\right)\right]^{1/2}}{\left[1+2\frac{\lambda(T)}{d}coth\left(\frac{t}{\lambda(T)}\right)\right]^{1/2}}\quad.
    \label{Eq_PPR_ft}
\end{equation}
where $t$ is the thickness of the two (assumed identical) superconducting samples and $f_0(0)$ is the resonant frequency at $T=0$. Here, we use the BCS s-wave temperature-dependence of the penetration depth in all fits. Note that the PPR technique measures the in-plane electromagnetic response of the superconducting films.\cite{Taber1990,Talanov2000}  It is possible  that the out-of-plane response could be different, as suggested by thermal conductivity measurements.\cite{Sood2016} \par



Boron doped diamond films (films "A","B", and "C") were deposited using the hot filament chemical vapor deposition (HFCVD) technique \cite{Kumar2017, Abdel2017, Kumar2018a}. The silicon substrate temperature was maintained at 850 \degree C and pressure of the chamber was $\sim$ 7 Torr. The gases used during the depositions were CH$_4$ (80 sccm), H$_2$ (3000 sccm) and B(CH$_3$)$_3$ with the flow rate maintained in such a way that the B/C ratio was approximately 10,000 ppm. The Hall concentration ($n_h$) and onset critical temperature ($T_{c,DC(Onset)}$) of film "A" were found to be $3.0 \times {10}^{21} {cm}^{-3}$ and 7.2 K, respectively. The film was deposited for 3 h and the thickness was found to be 1.5 $\mu m$ by cross-sectional electron microscopy. The other three films, covering a range of $T_c$ values, were grown under similar conditions with the full list of properties listed in Table 1 of the the Supplemental Material.

\begin{figure}[t!]
\includegraphics[width=0.5\textwidth]{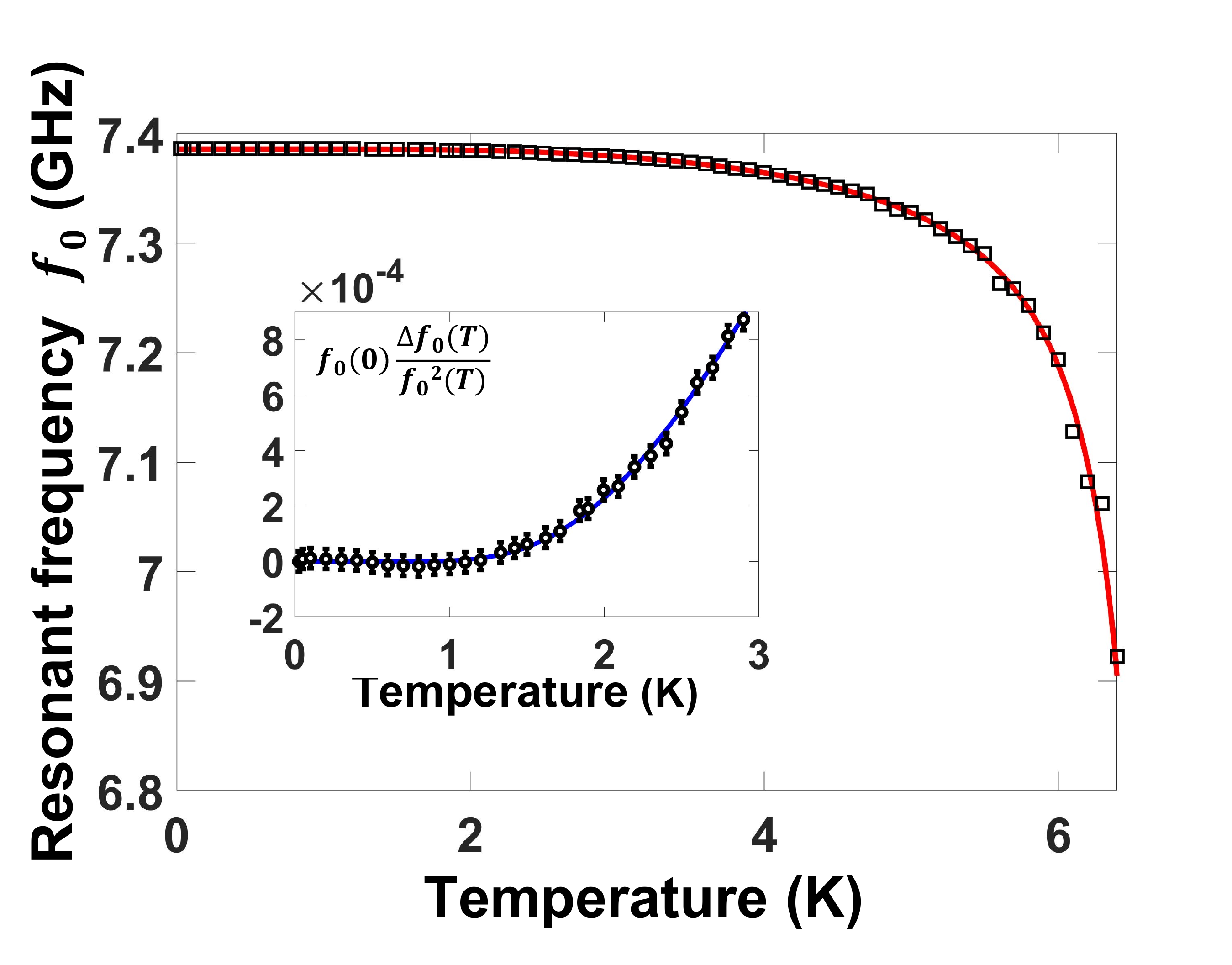}
\caption{ Temperature dependence of the resonant frequency $f_{0}(T)$ extracted from PPR measurement on Boron-doped diamond film "A" (black squares) and fits to the data using Eq.(\ref{Eq_PPR_ft}) (solid red line). Here the error bars are smaller than the symbol size. Inset shows the quantity $f_{0}(0)\Delta f_{0}(T)/f_{0}^2(T)$ vs temperature, which is proportional to $\Delta\lambda(T)/\lambda(0)$. Solid blue line is the fit to the data using Eq.(\ref{Eq_PPR_lowTlambda}) with $\Delta(0)=924.38\pm76.60\mu eV$. \label{Fig_SampleA}}
\end{figure}

The surface impedance measurements were done by means of the PPR technique.\cite{Taber1990,Pambianchi1994,Talanov2000} Two nominally identical films approximately $8\times6\;mm^2$, prepared by dicing a single sample, are placed face-to-face and sandwiching a Sapphire dielectric spacer of thickness $d=75\;\mu m$ or $430\;\mu m$ (see inset of Fig. \ref{Fig_SampleB}). The PPR is placed in a metallic enclosure and cooled to cryogenic temperatures. Two microwave coaxial cables make capacitive coupling to the PPR and induce a series of microwave resonances, and the complex transmitted signal $S_{21}$ vs. frequency is measured from $T=25\;mK$ to near $T_c$. The measurements are performed under vacuum in a BlueFors dilution refrigerator and the PPR assembly is firmly anchored to the mixing chamber plate. The measurements are repeated at several microwave powers and only the power-independent data is considered to exclude the influence of nonlinearity and microwave heating. These data are analyzed using the "phase versus frequency fit" method described in Ref. \cite{Petersan1998} to extract the resonant frequencies $f_0$ and quality factors $Q$ of the PPR modes. The $95\%$ confidence intervals for these two fitting parameters were used as their respective fitting uncertainties. The temperature of the PPR is systematically varied and $Q(T)$ and $f_0 (T)$ of the resonator are measured. As discussed below, these quantities are converted to the complex surface impedance, from which the surface resistance and magnetic penetration depth of the film can be determined.\cite{Taber1990,Pambianchi1994,Talanov2000} \par

The resonant frequency data for a PPR mode at frequency $f\simeq7.39\;GHz$ measured on film "A" is shown in Fig.\ref{Fig_SampleA}. A similar PPR resonant frequency data measured at $f\simeq8.08\;GHz$ measured on film "B" is shown in Fig.\ref{Fig_SampleB}. The first thing to note is the unusually large range of frequency shift of the modes, on the order of several hundred MHz. The fractional frequency shift is 5 times larger than previously published data.\cite{Taber1990}  Given the large dielectric spacer thickness this implies that the change in the penetration depth $\Delta\lambda(T)$ is unusually large for this material. \par

Equation (\ref{Eq_PPR_ft}) is fit to the $f_{0}(T)$ data (Fig. \ref{Fig_SampleA}) using the LMFIT function in Pythons's SciPy library \cite{LMfit}, where the 'basinhopping' optimization method \cite{bashhopp} was utilized. The values of $t$ and $d$ were fixed while $f_0(0)$, $\lambda(0)$ and $T_c$ were treated as free variables. An estimate of the zero temperature penetration depth $\lambda(0)=2.189\pm0.006\;\mu m$ and the critical temperature $T_c=6.717\pm0.001\;K$ for film "A" was extracted , where the errors correspond to $68\%$ confidence intervals. Note that it is often observed that the $T_c$ obtained from fitting to microwave frequency shift data is lower than the zero resistance $T_{c,DC}$ value,\cite{Anlage1989a} especially in granular materials.\cite{Muller1980,Ramachandran1992}  The same procedure is repeated for the other films and the extracted parameters are summarized in Table \ref{samples_tab}. \par

An order of magnitude estimate for the zero temperature penetration depth for a \textit{homogeneous} Boron-doped diamond material can be obtained using the London screening length $\lambda_L=\sqrt{\frac{m}{\mu_0 n e^2}}=97.1 nm$ where $n=n_h=3.0\times10^{21} cm^{-3}$ is the doping density of sample "A" and $e$, $m$ are the un-renormalized electron charge and mass. Estimates of $\lambda_L$ by other authors are on the same order of magnitude,\cite{Winzer05,Ort06} and the measured screening length $\lambda(0)$ is comparable to that estimated with the Ferrell-Glover-Tinkham sum rule at sub-THz frequencies ($\lambda \simeq 1 \mu$m).\cite{Ort06} The measured penetration depth is much larger than $\lambda_L$ hence it's value can not be explained by low carrier density alone and the effects of granularity should be considered.

The low temperature penetration depth $\lambda(T)$ of a homogeneous s-wave superconductor has an exponential dependence on temperature \cite{Prozorov2006}:
\begin{equation}
    \frac{\Delta\lambda(T)}{\lambda(0)}=\sqrt{\frac{\pi\Delta(0)}{2k_BT}}exp\left({-\frac{\Delta(0)}{k_BT}}\right),
    \label{Eq_PPR_lowTlambda}
\end{equation}
where $\Delta(0)$ is the zero temperature BCS superconducting gap. The inset of Fig.\ref{Fig_SampleA} shows the low temperature ($T<T_c/3$) behaviour of the frequency shift for sample "A". Here the change in penetration depth $\Delta\lambda(T)/\lambda(0) \propto \Delta f_{0}(T)/f_{0}^2(T)$ with $\Delta f_{0}(T) \equiv f_{0}(T)-f_{0}(0)$ is plotted. Our data indicates that $\Delta \lambda(T)$ is exponentially activated in temperature, consistent with the full energy gap expected for an s-wave superconductor. This indicates that granularity has not introduced sub-gap states into the superconductor. \par

Eqs.(\ref{Eq_PPR_ft}) and (\ref{Eq_PPR_lowTlambda}) were fit to the low temperature portion of the measured frequency shift (see solid blue line in the inset of Fig.\ref{Fig_SampleA}) with the value of $\lambda(0)=2.189\;\mu m$ fixed and $\Delta(0)$ as a fitting parameter, which yielded $\Delta(0)=924.38\pm76.60\mu eV$ , where the error reflects the $95\%$ confidence interval. Using $T_c=6.717K$ extracted from the full temperature fit to the same data, the superconducting gap to critical temperature ratio is calculated to be $\Delta(0)/k_BT_c=1.597$ somewhat less than 1.764 which is expected from weak-coupled BCS theory. However we believe that our film is a heavily doped granular material, and not a doped single crystal. Previous work on granular Boron-doped diamond films by scanning tunneling spectroscopy showed a range of values for the gap ratio ($0.63<\frac{\Delta(0)}{k_BT_c}<1.54$) \cite{Dahlem2010}. The authors of that work attributed this variability to an inverse proximity effect whereby the granular aspect of the sample created a range of gap values at the nanoscale. An estimate of the gap, $\Delta(0)/k_BT_c=1.5 \pm 0.25$, was previously obtained by gap spectroscopy.\cite{Ort06}

\begin{table*}
\caption{\label{samples_tab} Summary of the properties extracted from the PPR measurement for all samples.}
\begin{ruledtabular}
\begin{tabular}{cc|ccccc|c}
 Sample & Film thickness $t$ & dielectric spacer & $f_0(0)$ & $\lambda(0)$ & $T_c(RF)$ & $R_{eff}(0)$\footnote{Upper limit estimate of resistive losses.} & $\lambda_{BCS}^{dirty}(0)$ \\ \hline
 A & $1.5\pm0.2\;\mu m$ & $430\pm25\mu m$ sapphire & 7.39 GHz $\pm$ 0.14 MHz& $2.19 \pm 0.01 \mu m$ & 6.72 K $\pm$ 1 mK& 64.22 $\pm$ 0.39 $m\Omega$ & $3.14\;\pm\;0.04\;\mu m$ \\
 B & $3.5\pm0.2\;\mu m$ & $75\pm5\mu m$ sapphire & 8.08 GHz $\pm$ 0.86 MHz& $3.81 \pm 0.03 \mu m$ & 2.21 K $\pm$ 2 mK & 127.16 $\pm$ 0.66 $m\Omega$\footnote{\label{note1}These results were obtained in cooldowns where Cryoperm magnetic shielding was utilized.} & $7.33\;\pm\;0.66\;\mu m$ \\ 
 C &  $4.0\pm0.2\;\mu m$ & $75\pm5\mu m$ sapphire & 6.02 GHz$\pm$ 0.47 MHz & $3.82 \pm 0.07 \mu m$ & 1.54 K $\pm$ 2 mK& 12.24 $\pm$ 0.11 $m\Omega$\footnoteref{note1} & $4.79\;\pm\;0.43\;\mu m$ \\ 
 A & $1.5\pm0.2\;\mu m$ & $75\pm5\mu m$ sapphire & 5.30 GHz $\pm$ 1.26 MHz \footnote{\label{note2}This mode is different from the $f_0(0)=7.39 GHz$ mode listed above. The extracted values for $\lambda(0)$ and $T_c$ depend on the mode. This suggests that the sample may be inhomogeneous since the two modes send rf current through different parts of the films.} & $4.02 \pm 0.02 \mu m$\footnoteref{note2} & 5.69 K  $\pm$ 1 mK\footnoteref{note2} & 1.37 $\pm$ 0.02 $m\Omega$\footnoteref{note1} & $3.42\;\pm\;0.04\;\mu m$  \\

\end{tabular}
\end{ruledtabular}
\end{table*}

The nonlinearity of the material can be quantified through the self-Kerr coefficient $K_{11}$, \cite{Maleeva2018} which is defined as the amount of resonant frequency shift per added photon. For this measurement, sample "B" was kept at a base temperature of $T=100mK$ and the resonance was measured while sweeping the excitation power. The circulating photon number in the resonator was roughly estimated as $\braket{n}\approx 4QP_{in}/(hf_0^2)$, where $P_{in}$ is the power entering the resonator, $Q$ is the loaded quality factor and $h$ is Planck's constant. A linear fit to the shift in resonance frequency vs the average circulating photon number data yields the slope of $K_{11}=15.25\,mHz/photon$. This value can be substantially enhanced by decreasing the thickness of the film.

The extracted zero temperature penetration depth values for the Boron-doped diamond films are significantly larger than the values reported for other conventional superconductors, which range between $50nm-500nm$. \cite{Greytak1964,Anlage1992,Prozorov2000,Hashimoto2009} A possible explanation for the large value of the penetration depth is that these films are highly granular, giving rise to a large effective screening length due to a combination of Meissner and Josephson screening. This is consistent with numerous previous studies showing granular behaviour in Boron-doped diamond films. \cite{Dubrovinskaia2006,Willems2010,Dahlem2010,Moshchalkov2013,Li2014,KLEMENCIC2021} \par

The effective penetration depth of polycrystalline and nanocrystalline superconductors can be estimated using the Laminar Model, \cite{Hylton1989Laminar} which predicts that the effective penetration depth of such superconductors strongly depends on the coupling strength between the grains, approaching the intrinsic value of the penetration depth of the grains in the strong coupling limit. As discussed in Supp. Mat. section II, for an MCD or NCD film in the small grain and weak coupling limit one finds an effective penetration depth of $\lambda_{eff} = \sqrt{\frac{\hbar \rho_{barrier}}{\pi \mu_0 \Delta}} \sqrt{\frac{t_{barrier}}{a}}$, where $t_{barrier}$ and $\rho_{barrier}$ are the thickness and resistivity of the barrier material between the grains, and $a$ is the grain size. This can be compared to the dirty limit BCS expression of $\lambda_{BCS}^{dirty} = \sqrt{\frac{\hbar \rho_n}{\pi \mu_0 \Delta}}$, where $\rho_n$ is the resistivity of the assumed-homogeneous film.\cite{Zmuidzinas2012}  Depending on the $t_{barrier}$ and $\rho_{barrier}$ parameter values (which are unknown), in the MCD and NCD cases one can find values of the effective screening length greater than, or less than, the standard BCS dirty limit value. Table I contains estimates of $\lambda_{BCS}^{dirty}$ for our films, and they are on the same order of magnitude of the measured screening length values, consistent with our estimate of $\lambda_{eff}$ \par

Similarly large values of penetration depth are observed in other amorphous and granular superconductors: $\lambda(0)=0.39 \mu m$ in NbN \cite{Pambianchi1994}, $\lambda(0)>0.51 \mu m$ in Mo$_3$Si \cite{Wordenweber}, $\lambda(0)=0.575 \mu m$ in TiN \cite{Vissers2010}, $\lambda(0)=0.645 \mu m$ in Mo-Ge \cite{MissertThesis1989}, $\lambda(0)>0.65 \mu m$ in Nb$_3$Ge \cite{Wordenweber} and $\lambda(0)=1.2 \mu m$ in granular aluminium films \cite{Cohen1968a}. Such large values of penetration depth make Boron-doped diamond films an appealing material to be used in applications where large inductance in needed, such as microwave kinetic inductance detectors (MKID) or superconducting microresonator bolometers \cite{Zmuidzinas2012}. \par

\begin{figure}[t!]
\centering
\includegraphics[width=0.5\textwidth]{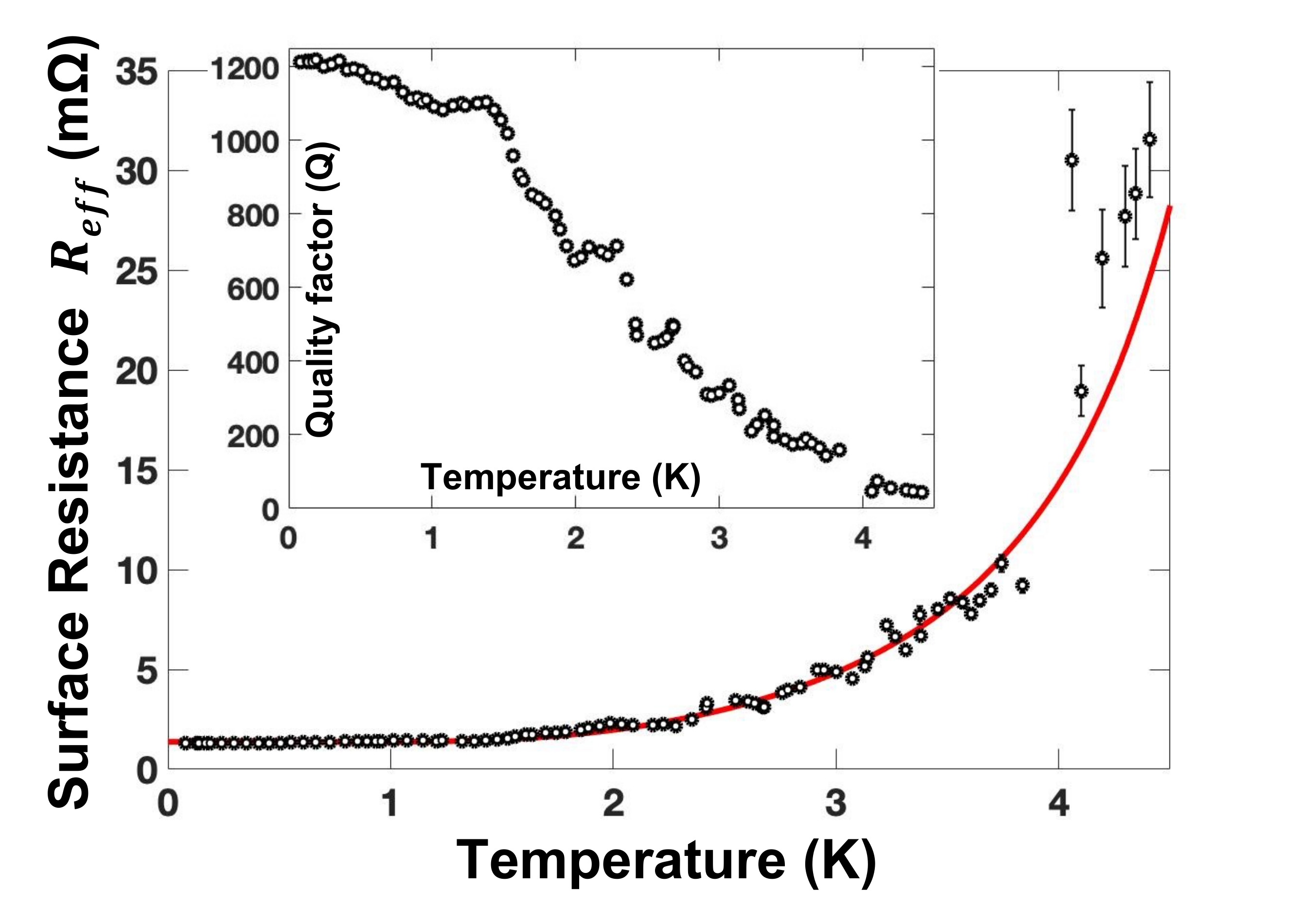}
\caption{Effective surface resistance $R_{eff}$ of the Boron-doped diamond film "A" measured at a frequency of $f=4.9-5.3$ GHz (black $\circ$). Also shown is the fit to the data (red line) using Eq. (\ref{Eq_BCS_Res}) with  $\lambda(0)=4.03\;\mu m$, $T_c=5.69\;K$ and $\Delta(0)=1.764k_BT_c$ as fixed parameters; and $R_N=696.77\pm46.21m\Omega$, $\epsilon_r=6.85\pm 0.24$ as fitting parameters. The inset shows the raw quality factor vs temperature data where the error bars are smaller than the symbol size.  The errorbars for $R_{eff}$ were obtained by means of standard error propagation as $\sigma_{R_{eff}}/R_{eff}(T)=\sqrt{\left(\sigma_{f}/f(T)\right)^2+\left(\sigma_{Q}/Q(T)\right)^2}$.}
\label{Rs_T}
\end{figure} 

The effective surface resistance of these films is obtained from the measured quality factor $Q(T)$ through $R_{eff}(T)=\pi\mu_0f_{0}(T)d/Q(T)$. \cite{Taber1990,Talanov2000}  Figure \ref{Rs_T} shows the temperature dependence of the effective surface resistance $R_{eff}(T)$ of the Boron-doped diamond film "A". In this case the film was measured using the PPR with a $d=75\;\mu m$ thick Sapphire dielectric spacer. The entire setup was shielded from ambient magnetic field using a cylindrical Cryoperm magnetic shield.\par

The $R_{eff}$ data shows substantial residual loss in the zero temperature limit, as summarized in Table \ref{samples_tab}. There are several contributions to the measured $Q$, including coupling loss, radiation loss from the sides of the PPR ($Q_{rad}>8\times10^5$ based on estimates from Ref. \cite{Taber1990}) and through the films, absorption in the dielectric spacer (expected to be small because $Q_{dielectric}=1/tan\delta \sim 10^6$ for sapphire), enhanced Ohmic loss due to the finite thickness of the films, and losses in the grain boundaries, \cite{Hylton1988} which have been observed to be rich in Boron \cite{Kumar2018} and trans-polyacetylene.\cite{Kumaran15}  Finally is the possibility of enhanced losses due to trapped magnetic flux in the films.\cite{Pambianchi1993,Pompeo2020}  A simplified fit to the effective surface resistance vs. temperature data can be obtained as follows. \par

The effective surface resistance temperature dependence mainly arises from an intrinsic BCS contribution ($R_{BCS}(T)$) and related extrinsic (finite thickness) contributions\cite{Klein1990}

\begin{equation}
R_{eff}(T)=R_{BCS}(T)\times{\cal F}(t/\lambda(T))+R_{trans},
\label{Eq_BCS_Res}
\end{equation}
\noindent where
\begin{eqnarray*}
\frac{R_{BCS}(T)}{R_{N}}=\sqrt{\frac{\sqrt{\left(\sigma_1/\sigma_n\right)^2+\left(\sigma_2/\sigma_n\right)^2}-\sigma_2/\sigma_n}{\left(\sigma_1/\sigma_n\right)^2+\left(\sigma_2/\sigma_n\right)^2}}\quad,\\
R_{trans}(T)=\frac{\sqrt{\epsilon_r}}{Z_0}\frac{\left(\mu_0\omega\lambda(T)\right)^2}{\left[sinh(t/\lambda(T))\right]^2}\quad,\\
{\cal F}(t/\lambda(T))=coth(t/\lambda(T))+\frac{t/\lambda(T)}{\left[sinh(t/\lambda(T))\right]^2}\quad.\nonumber
\end{eqnarray*}

\noindent Here $R_{BCS}(T)$ is the bulk Meissner state BCS surface resistance, \cite{Turneaure1991,Zmuidzinas2012,Gurevich2017}  where we numerically solved for $\frac{\sigma_1(\omega,T)}{\sigma_N}$ and $\frac{\sigma_2(\omega,T)}{\sigma_N}$ using Mattis-Bardeen theory, \cite{Mattis1958} ${\cal F}$ is an enhancement factor due to the finite thickness of the film, \cite{Klein1990} $R_{trans}$ accounts for the radiation loss into the substrate of effective permittivity $\epsilon_r$, and $Z_0=377 \Omega$ is the impedance of free space. Equation (\ref{Eq_BCS_Res}) was fit to the effective surface resistance data (red line in Fig. \ref{Rs_T}) with fixed values of $\lambda(0)=4.03\;\mu m$, $T_c=5.69\;K$, $\Delta(0)=1.764k_BT_c$, which were extracted from a fit of the associated $f_0(T)$ data near 5.3 GHz to  Eq. (\ref{Eq_PPR_ft}). The normal state surface resistance of the Boron-doped diamond $R_N$ and the effective permittivity of the substrate $\epsilon_r$ were used as the fitting parameters. The most accurate fit to the data was achieved with $R_N=696.77\pm46.21m\Omega$, and $\epsilon_r=6.85\pm 0.24$.  The effective dielectric constant $\epsilon_r$ is intermediate in value between that of diamond and silicon, consistent with a diamond-seeded silicon substrate (see Supp. Mat. Section I).  We note that the normal state surface resistance is related to the normal state resistivity as $R_N=\sqrt{\mu_0 \omega \rho_N/2}$.  The resulting estimate of $\rho_N=2.32\;\pm0.31m\Omega-cm$ is close to the independently measured value of $\rho_{N,DC}=5.5\;\pm0.1\;m\Omega-cm$, giving confidence in this fit. The losses associated with the other extrinsic contributions such as radiation loss out the sides of the PPR and the dielectric loss are small compared to the contributions calculated here.\cite{Taber1990,Pambianchi1994,Talanov2000} \par

Boron-doped diamond has also been proposed as a platform to build future hybrid quantum devices \cite{Coleman2020}, thanks to its unique properties. By controlling the dopant type and concentration in Diamond films one can create an insulator, a semiconductor (both p and n type), an optically-transparent electrode,\cite{Wachter2016} or a superconductor, all using the same starting material. Along these lines, Watanabe \textit{et al.} demonstrated  very uniform vertical SNS Josephson junctions solely with BDD by carefully controlling the doping concentration during the deposition of the thin-film \cite{Watanabe2012}. This, and recent successful demonstrations of nanosize and microsize patterning processes for Boron-doped diamond films, \cite{Kageura2018} indicates that the prospect of building complex superconducting devices using Boron-doped diamond films is possible. Finally we note that quantum devices based on nitrogen vacancy centers in diamond offer new opportunities for superconducting quantum devices.\cite{Kubo11}

In this work we present the measurements of in-plane complex surface impedance of Boron-doped diamond films using a microwave resonant technique. We demonstrate that these samples have very large penetration depth which is consistent with the local-limit estimates that are appropriate for granular superconductors. Furthermore, we provide additional evidence that Boron-doped diamond is an s-wave superconductor with few sub-gap states and shows promise for high kinetic inductance applications.

\section*{Supplementary Material}
See supplementary material for the detailed film microstructure and discussion about the screening length in granular superconductors.

\section*{Acknowledgments}
We thank Chung-Yang Wang for assistance with PPR measurements.  B.O. and S.M.A. acknowledge support from the U.S. Department of Energy/ High Energy Physics through Grant No.DESC0017931 (support for B.O.) and DOE/Basic Energy Sciences through Grant No.DESC0018788 (measurements). M.S.R. and D.K. would like to acknowledge the funding and extended supported from the Department of Science and Technology (DST), New Delhi, that facilitated the establishment of “Nano Functional Materials Technology Centre” (Grant $\#$s: SRNM/NAT/02-2005, DST/NM/JIIT-01/2016 (G); SR/NM/NT-01/2016).

\section*{Data Availability}
The data that support the findings of this study are available from the corresponding author upon reasonable request.

\bibliography{MendeleyLibrary}

\end{document}